\documentclass[aps,pre,twocolumn,groupedaddress]{revtex4-1}

\usepackage{xcolor,hyperref}
\hypersetup{
   colorlinks,
   linkcolor={blue!50!black},
   citecolor={blue!50!black},
   urlcolor={blue!80!black}
}

\usepackage{amsmath}
\usepackage{amssymb}
\usepackage{color}

\usepackage{graphicx}

\DeclareMathAlphabet{\mathitbf}{OML}{cmm}{b}{it}

\newcommand{\zv}{\mathitbf z}
\newcommand{\xv}{\mathitbf x}
\newcommand{\dv}{\mathitbf d}

\newcommand{\calBold}[1]{\mbox{\boldmath${\cal #1}$}}
\newcommand{\mathBold}[1]{\mbox{\boldmath$#1$}}

\begin{document}

\title{Finite-size effects in the nonphononic density of states in computer glasses}



\author{Edan Lerner}
\email{e.lerner@uva.nl}
\affiliation{\vspace{0.15cm}Institute for Theoretical Physics, University of Amsterdam, Science Park 904, Amsterdam, Netherlands}

\begin{abstract}
The universal form of the density of nonphononic, quasilocalized vibrational modes of frequency $\omega$ in structural glasses, ${\cal D}(\omega)$, was predicted theoretically decades ago, but only recently revealed in numerical simulations. In particular, it has been recently established that, in generic computer glasses, ${\cal D}(\omega)$ increases from zero frequency as $\omega^4$, independent of spatial dimension and of microscopic details. However, in [E. Lerner, and E. Bouchbinder, Phys.~Rev.~E \textbf{96}, 020104(R) (2017)] it was shown that the preparation protocol employed to create glassy samples may affect the form of their resulting ${\cal D}(\omega)$: glassy samples rapidly quenched from high temperature liquid states were shown to feature ${\cal D}(\omega)\!\sim\!\omega^\beta$ with $\beta\!<\!4$, presumably limiting the degree of universality of the $\omega^4$ law. Here we show that exponents $\beta\!<\!4$ are only seen in small glassy samples quenched from high-temperature liquid states --- whose sizes are comparable to or smaller than the size of the disordered core of soft quasilocalized vibrations --- while larger glassy samples made with the \emph{same protocol} feature the universal $\omega^4$ law. Our results demonstrate that observations of $\beta\!<\!4$ in the nonphononic density of states stem from finite-size effects, and we thus conclude that the $\omega^4$ law should be featured by any sufficiently-large glass quenched from a melt.
\end{abstract}


\maketitle

\section{introduction}
The form of low-frequency spectra of glassy solids remains a subject of lively debates, despite decades of theoretical \cite{soft_potential_model_1991,Schober_prb_1992,Gurevich2003,matthieu_PRE_2005,Schirmacher_prl_2007,eric_boson_peak_emt,silvio}, numerical \cite{Schober_Laird_numerics_PRL,Schober_Oligschleger_numerics_PRB,ohern2003,barrat_3d,vincenzo_epl_2010,modes_prl,modes_prl_2018,ikeda_pnas} and experimental research \cite{Effect_of_Densification_prl_2006,experiments_300K_vSiO2,experiments_1620K_vSiO2,Monaco_prl_2011}. Low-frequency excitations are believed to be responsible for several important glassy phenomena, ranging from yielding \cite{argon_simulations,pine_emulsions_stzs,barrat_yielding_jcp_2004,Schall1895} and shear banding \cite{Ozawa6656,falk_shear_banding_pre_2018,MW_yielding_2018_pre,Sylvain_pre_rejuvenation} to anomalous wave attenuation \cite{scattering_jcp}. Relations between low-frequency excitations and both relaxation in supercooled liquids \cite{schober1993_numerics,Schober_correlate_modes_dynamics,widmer2008irreversible}, and low-temperature thermodynamics of glassy solids \cite{Anderson,Phillips,khomenko2019depletion}, have also been previously suggested. 

Numerical investigations of computer glass models have contibuted significantly to our understanding of many problems in glass physics, and, in particular, to revealing the structural and statistical properties of soft, quasilocalized modes (QLMs) in glassy solids. Pioneered by Schober and coworkers decades ago \cite{Schober_Laird_numerics_PRL,Schober_Oligschleger_numerics_PRB}, computational investigations have revealed that the lowest-frequency vibrational modes in computer glasses are quasilocalized: they consist of a disordered core, decorated by an algebraically-decaying field away from the core \cite{modes_prl,atsushi_core_size_pre,pinching_2019}. Furthermore, it was discovered that the distribution ${\cal D}(\omega)$ of these nonphononic modes follows a universal $\sim\!\omega^4$ law (with $\omega$ denoting the frequency), independent of spatial dimension \cite{modes_prl_2018}, or microscopic details \cite{modes_prl,ikeda_pnas}. Remarkably, these findings are consistent with theoretical predictions put forward some time ago \cite{soft_potential_model_1991,Schober_prb_1992,Gurevich2003}. 

Universality of the nonphononic ${\cal D}(\omega)\!\sim\!\omega^4$ density of states with respect to glass preparation protocol was also reported \cite{protocol_prerc,cge_paper,LB_modes_2019}, however with some caveats; in particular, in \cite{protocol_prerc} it was shown that glassy samples created by an infinitely-fast quench from high-temperature liquid states, or by continuous quenches at very high rates, feature ${\cal D}(\omega)\!\sim\!\omega^\beta$ with $\beta\!<\!4$. In \cite{protocol_prerc} it was further suggested that the observation of $\beta\!<\!4$ could be consistent with the Soft Potential Model \cite{soft_potential_model_1991,Schober_prb_1992}, which states that very unstable states --- presumably similar to those obtained by instantaneous \emph{overdamped} quenches of high temperature liquid states --- could feature $\beta\!<\!4$.

In this paper we show that the observations of ${\cal D}(\omega)\!\sim\!\omega^\beta$ with $\beta\!<\!4$ made in \cite{protocol_prerc} are, in fact, the result of too-small glassy samples employed. The results presented here for ${\cal D}(\omega)$ were obtained using the same computer glass model as employed in \cite{protocol_prerc}. We prepared large ensembles of glassy samples by instantaneous quenches from high-temperature liquids states, and by gradually varying the system size $N$, we show in what follows that the exponent $\beta$ that characterizes the distribution of soft, nonphononic QLMs varies from $\beta\!<\!4$ in small systems, to $\beta\!=\!4$ in larger systems. We further show that the characteristic length $\xi_g$ of soft QLMs --- recently shown to be well-represented by the spatial form of linear responses to localized force dipoles \cite{pinching_2019} --- is comparable to the system sizes in which $\beta\!<\!4$ is observed, suggesting that finite-size effects in ${\cal D}(\omega)$ depend on the ratio $\xi_g/L$, with $L$ denoting the linear system size. We expect our results to serve as constraints on the formulation of future theories about the statistical mechanics of meso-scale elasticity in glassy solids. 

\begin{figure}[!ht]
\centering
\includegraphics[width = 0.4\textwidth]{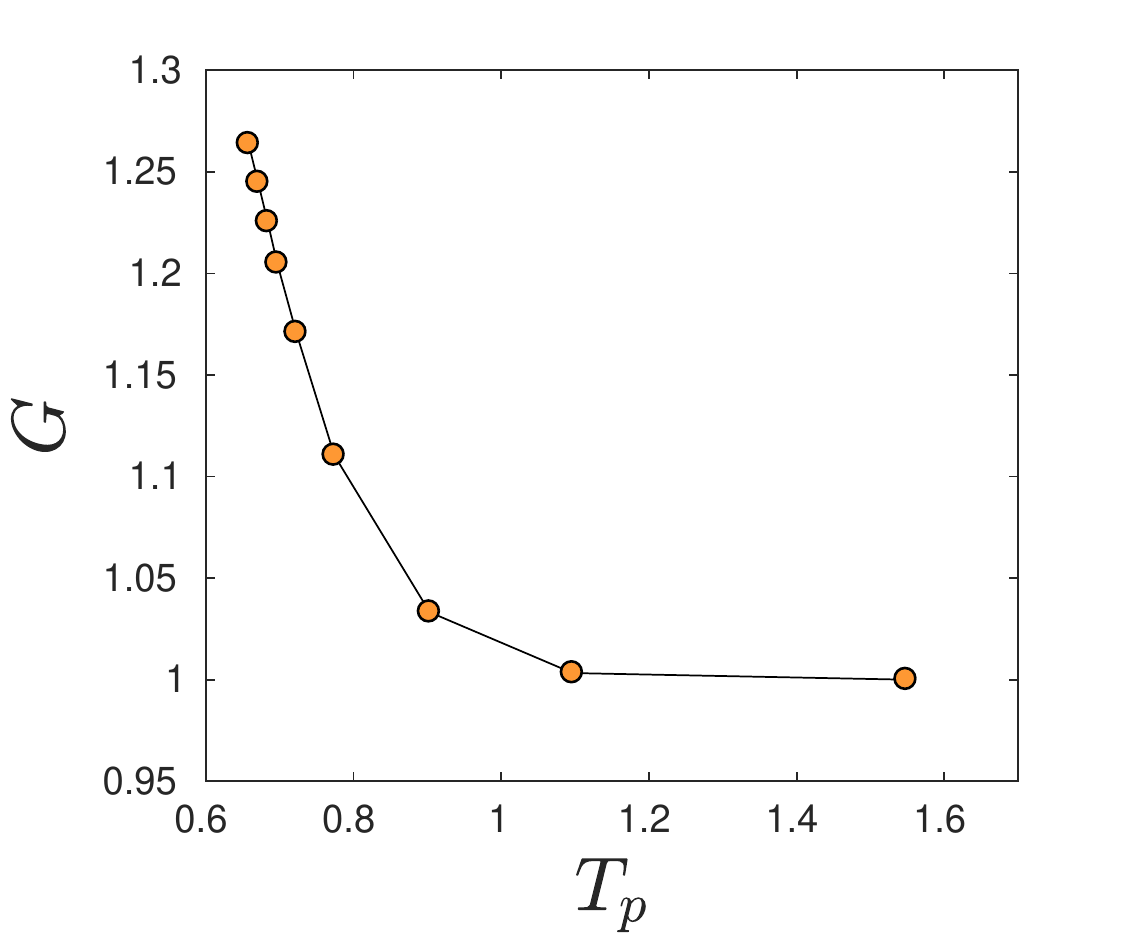}
\caption{\footnotesize Sample-to-sample mean athermal shear modulus $G$, plotted against the equilibrium parent temperature $T_p$ from which our glassy samples were instantaneously quenched. In this work we express temperatures in terms of the onset temperature of the high-$T_p$ plateau of $G$, and elastic moduli in terms of the high-$T_p$ plateau of $G$ itself.}
\label{fig:G_fig}
\end{figure}

\section{Models, methods and units}
The model employed here is described in detail in e.g.~\cite{cge_paper}; it is a binary mixture of `large' and `small' particles interacting via a radially-symmetric, purely repulsive $\!\propto\! r^{-10}$ pairwise potential, with $r$ denoting the distance between pairs of particles. In what follows, we refer to the \emph{parent temperature} $T_p$ of an ensemble of glassy samples as the equilibrium temperature from which they were instantaneously quenched, see e.g.~\cite{cge_paper,pinching_2019}. Parent temperatures $T_p$ are expressed in terms of the onset parent temperature $T_{\mbox{\tiny onset}}$ above which the sample-to-sample mean of the athermal shear modulus $G$ \cite{cge_paper} of underlying inherent states plateaus, as shown in Fig.~\ref{fig:G_fig}. We express lengths in terms of $a_0\!\equiv\!(V/N)^{1/3}$ where $V\!=\! L^3$ is the volume. The mass $m$ is the same for all particles, and is chosen as our microscopic units of mass. Elastic moduli are expressed in terms of the high-parent-temperature plateau $G_\infty$ of the sample-to-sample mean athermal shear modulus (as seen in Fig.~\ref{fig:G_fig}), frequencies in terms of $c_\infty/a_0$ where $c_\infty\!\equiv\!\sqrt{G_\infty/\rho}$ represents the speed of sound of glasses quenched from $T_p\!>\! T_{\mbox{\tiny onset}}$, and $\rho\!=\! mN/V$ is the mass density. 

Glassy samples were made by first equilibrating independent liquid states at a very high temperature of $T\!=\!2.5\!>\! T_{\mbox{\tiny onset}}$, and then instantaneously quenching those states to zero temperature using a standard conjugate gradient algorithm. We prepared 320000, 80000, 20000, 5000, and 1250 glassy samples of $N=2048,8192,32768, 131072$ and 524288 particles, respectively. Ensemble sizes were chosen such that the quality of the statistics is independent of system size.

\begin{figure*}[!ht]
\centering
\includegraphics[width = 1.00\textwidth]{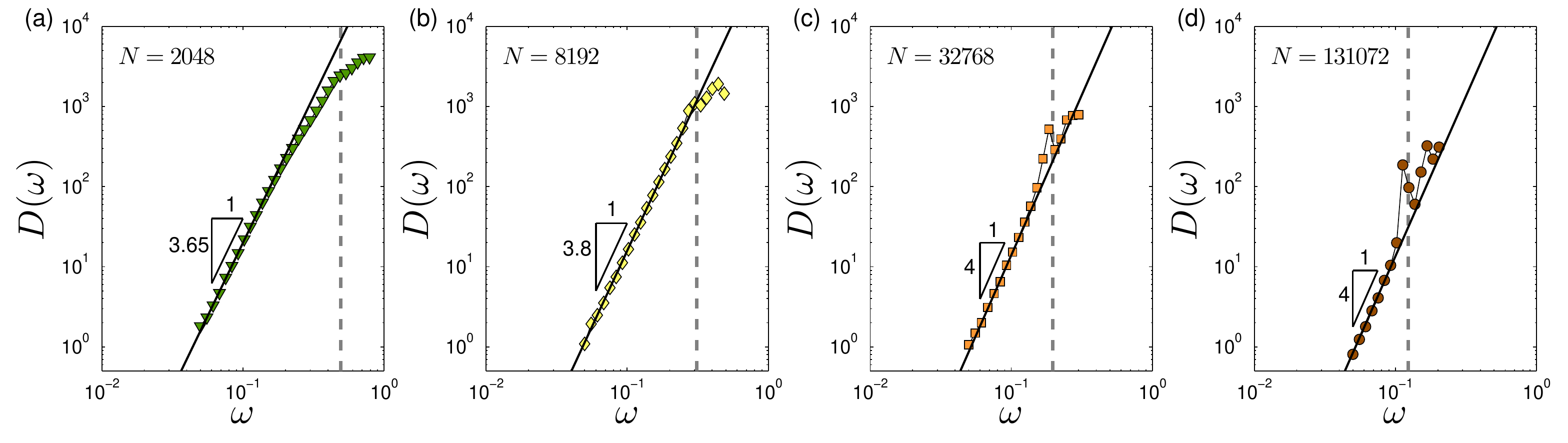}
\caption{\footnotesize The low-frequency regime of the total density of states $D(\omega)$, measured in our ensembles of glassy samples quenched from high temperatures, of (a) $N\!=\!2048$, (b) $N\!=\!8192$, (c) $N\!=\!32768$, and (d) $N\!=\!131072$ particles. The vertical dashed lines mark the first phonon frequency $2\pi c_\infty/L$. These data show that $D(\omega)\!\sim\!\omega^\beta$ with $\beta\!<\!4$ is a finite-size effect.}
\label{fig:dos_fig}
\end{figure*}

In what follows we discuss the statistical properties of vibrational frequencies $\omega$ associated with vibrational modes $\mathBold{\psi}_\omega$, which are solutions to the eigenvalue problem 
\vspace{-0.4cm}
\begin{equation}
\calBold{M}\cdot\mathBold{\psi}_\omega = m\omega^2\mathBold{\psi}_\omega\,,
\end{equation}
where $\calBold{M}\!\equiv\!\frac{\partial^2U}{\partial\xv\partial\xv}$ is the Hessian of the potential $U(\xv)$ evaluated at a local minimum, and $\xv$ denotes the particles' coordinates. For each member of our ensembles of glassy samples we calculated the lowest 100 vibrational frequencies using ARPACK \cite{arpack}. A discussion regarding the conditions under which ${\cal D}(\omega)$ can be cleanly observed, without hindrance by hybridizations with phonons, can be found in \cite{phonon_widths}.

\begin{figure}[!ht]
\centering
\includegraphics[width = 0.47\textwidth]{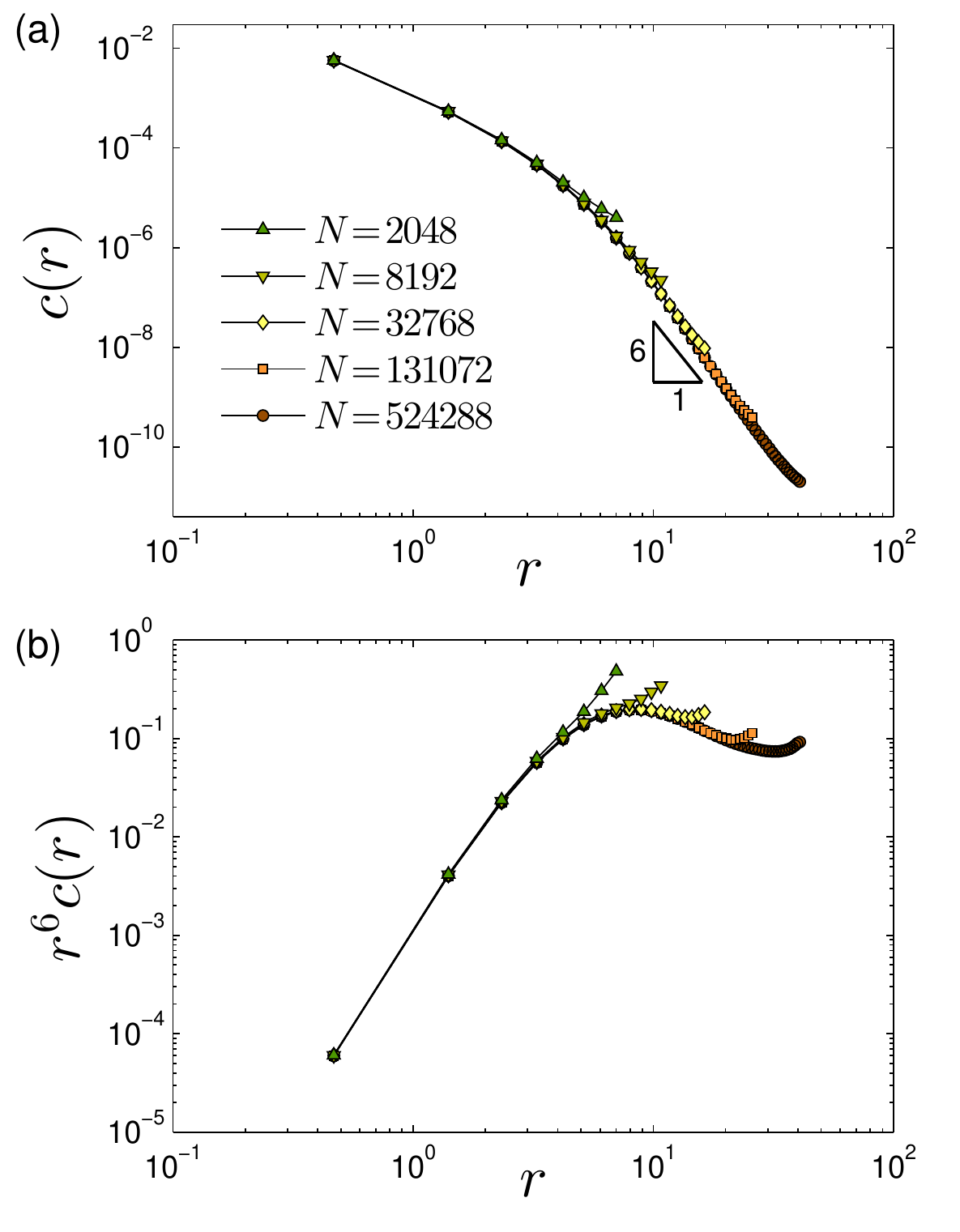}
\caption{\footnotesize (a) Decay functions $c(r)$ calculated on the responses to local force dipoles, as defined in the text, plotted against the distance $r$ from the applied force dipole, for different system sizes as reported in the legend. We expect to see a crossover to the continuum $r^{-6}$ scaling at $r\!\gtrsim\!\xi_g$, with $\xi_g$ representing the QLMs' core size. (b) Scaling the decay functions by $r^6$ shows that the smallest systems employed here are too small to exhibit the crossover to the expected continuum scaling.} 
\label{fig:dipole_decay}
\end{figure}

\section{Results}

As mentioned above, in \cite{protocol_prerc} it was shown that glassy samples quenched instantaneously from high parent temperatures $T_p$ feature a density of nonphononic quasilocalized modes ${\cal D}(\omega)\!\sim\!\omega^\beta$ with $\beta\!<\!4$. In Fig.~\ref{fig:dos_fig} we show the low-frequency regime of the \emph{total} density of states $D(\omega)$ measured in our glassy samples. Importantly, we note that below the lowest phonon frequency $2\pi c_\infty/L$, the nonphononic and total density of states are the same, i.e. $D(\omega)\!=\!{\cal D}(\omega)$. Consistent with \cite{protocol_prerc}, we find $\beta\!\approx\!3.65$ for $N\!=\!2048$; however, upon increasing the system size we find $\beta\!\approx\!3.8$ for $N\!=\!8192$, and $\beta\!\approx\!4$ for $N\!\ge\!32768$, in agreement with previous results \cite{cge_paper,SciPost2016,ikeda_pnas}. 

What gives rise to this finite-size effect in the exponent~$\beta$? We assert below that the finite-size effect seen in the exponent $\beta$ is the result of soft QLMs not `fitting' well in the simulation box. To establish this, we build on the recent result \cite{pinching_2019} that the length that characterizes the linear response to local force dipoles agrees well with the core size of QLMs. We thus first aim at measuring the length that characterizes the response to local force dipoles in our glassy samples.

To this aim, we carry out the following procedure, also employed in \cite{pinching_2019}; we randomly select pairs $ij$ of weakly interacting particles, and calculate the response $\zv^{(ij)}$ as
\begin{equation}
\zv^{(ij)} = \calBold{M}^{-1}\cdot\dv^{(ij)}\,,
\end{equation}
where the (dimensionless) dipole force $\dv^{(ij)}$ is defined as
\begin{equation}
\dv^{(ij)} \equiv \frac{\partial r_{ij}}{\partial\xv}\,,
\end{equation}
and $r_{ij}$ is the pairwise distance between particles $i$ and $j$. We normalized the fields $\zv^{(ij)}$, namely $\hat{\zv}^{(ij)}\!=\!\zv^{(ij)}/|\zv^{(ij)}|$, and, for each other pair of particles $lq\!\ne\! ij$, we calculate the projection $\hat{\zv}^{(ij)}\!\cdot\!\dv^{(lq)}$. Finally, we define the decay function $c(r)$ as
\begin{equation}
c(r) \equiv \big< \underset{r_{ij,lq}\approx r}{\mbox{median}_{lq}}\big[ (\hat{\zv}^{(ij)}\!\cdot\!\dv^{(lq)})^2 \big] \big>_{ij}\,.
\end{equation}
where $r_{ij,lq}$ stands for the distance between the average position of the pair $ij$ and the average position of the pair $lq$, and the angular brackets stand for an average over different pairs $ij$ to which a force dipole is applied, and over different glassy samples. We employ the median over pairs $lq$ that are at a distance $\approx\! r$ from the pair $ij$ --- rather than the mean --- in order to suppress large fluctuations that can occur in the projections \cite{breakdown}. We further note that the choice to only apply force dipoles to weakly interacting pairs is motivated in \cite{pinching_2019}, where it was shown that this strategy produced responses that are generally softer, and thus better representatives of soft QLMs.

Our results are shown in Fig.~\ref{fig:dipole_decay}; in panel (a) we show the raw decay functions $c(r)$ measured as explained above, where the different symbols represent the different system sizes considered. Since $c(r)$ scales as the strain field squared, we expect that at large distances $r$ from the applied force dipole, $c(r)$ should scale as $r^{-6}$ (in 3D). In order to better visualize the characteristic decay length, we plot in panel (b) the product $r^6c(r)$. This representation of the decay functions reveals that the crossover to the expected continuum behavior is \emph{not seen at all} for $N\!=\!2048$, and only initial signs of this crossover are seen for $N\!=\!8192$. We note that the uprise seen at the very last points of each signal is a result of the periodic boundary conditions employed. Finally, the shape of $c(r)$ depends on several details, in particular the distance to the unjamming transition \cite{ohern2003,liu_review,van_hecke_review} as discussed and shown in \cite{breakdown,robbie_auxetic}. 

The signals for $N\!>\!8192$ indicate that the crossover lengthscale to the expected continuum behavior --- estimated as the maximum of the bump in $r^6c(r)$ --- is about 8 particle diameters long, indicating that the effective diameter of QLMs is $\approx\!16$ particle diameters long (twice the effective radius as extracted from $c(r)$). Clearly, for $N\!\lesssim\!16^3$ QLMs cannot comfortably fit in the glassy solid. We propose that this explains why finite-size effects in ${\cal D}(\omega)$ are observed in these smaller system sizes.

\section{Summary, discussion and outlook}

In this work we have shown that generic computer glasses quenched quickly from high-parent-temperature liquid states feature the universal exponent $\beta\!=\!4$ in their nonphononic density of states ${\cal D}(\omega)\!\sim\!\omega^\beta$ --- if large-enough glassy samples are considered ---, and that smaller exponents $\beta\!<\!4$ result from a finite-size effect. We further showed that $\beta\!<\!4$ is seen when the system size is comparable to the core size of QLMs (represented by the response to local force dipoles \cite{pinching_2019}). Our findings correct the misinterpretation of numerical results presented and discussed in \cite{protocol_prerc}, where it was incorrectly concluded that exponents $\beta\!<\!4$ are not a finite-size effect, but rather have to do with the physical process of glass formation by rapid quenches from high-temperature liquid states. Our results also cast doubts on the claims made in \cite{Paoluzzi_2019}, where the change in $\beta$ induced by deeper supercooling was attributed to the suppression of phononic (plane-wave) modes.  

Our findings strongly suggest that the universality of the $\omega^4$ law extends to glasses formed from \emph{any} parent temperature, conditioned that those glasses are sufficiently larger than the core size of QLMs. Importantly, in \cite{pinching_2019} it was shown that the core size of QLMs decreases with decreasing parent temperature, explaining why the $\omega^4$ law of the nonphononic density of states can be measured in smaller glassy samples quenched from low parent temperatures, or quenched at slow rates into the glassy~phase.

Further support for the physical picture proposed in this work was put forward in \cite{atsushi_core_size_pre}, where it was shown that the far, algebraically-decaying field of QLMs stabilizes their energetically-unstable core. It is therefore reasonable to expect that if QLMs are deprived from their far fields --- as expected to occur in small glassy samples --- those QLMs will be less stable and have lower frequencies, possibly leading to $\beta\!<\!4$ in their nonphononic density of states. 

Finally, we propose that the finite-size effect reported here, in which a system constrains itself by virtue of its small size, giving rise to reduced stability manifested by the appearance of soft modes, might be an instance of a more general phenomenon. For example, in \cite{nonlinear_excitations} it was reported that small packings of frictionless hard spheres possess relatively fewer small gaps between nearby particles that are \emph{not} in contact, compared to the abundance of small gaps in larger packings. Since smaller gaps are known to stabilize hard-sphere packings \cite{nonlinear_excitations}, that phenomenon bears similarities to the reduced stability that small computer glasses possess, as reported here. 

In light of the findings reported here, it seems important to explore which additional finite-size effects on QLMs appear in computer glasses, and how those effects influence our insights and conclusions regarding the properties of macroscopic structural glasses. 

\acknowledgements
Warm thanks to Eran Bouchbinder for exhilarating discussions. We further thank Geert Kapteijns, Corrado Rainone, and Karina Gonzalez Lopez for their comments on the manuscript. E.~L.~acknowledges support from the Netherlands Organisation for Scientific Research (NWO) (Vidi grant no.~680-47-554/3259).


\begin{thebibliography}{48}%
\makeatletter
\providecommand \@ifxundefined [1]{%
 \@ifx{#1\undefined}
}%
\providecommand \@ifnum [1]{%
 \ifnum #1\expandafter \@firstoftwo
 \else \expandafter \@secondoftwo
 \fi
}%
\providecommand \@ifx [1]{%
 \ifx #1\expandafter \@firstoftwo
 \else \expandafter \@secondoftwo
 \fi
}%
\providecommand \natexlab [1]{#1}%
\providecommand \enquote  [1]{``#1''}%
\providecommand \bibnamefont  [1]{#1}%
\providecommand \bibfnamefont [1]{#1}%
\providecommand \citenamefont [1]{#1}%
\providecommand \href@noop [0]{\@secondoftwo}%
\providecommand \href [0]{\begingroup \@sanitize@url \@href}%
\providecommand \@href[1]{\@@startlink{#1}\@@href}%
\providecommand \@@href[1]{\endgroup#1\@@endlink}%
\providecommand \@sanitize@url [0]{\catcode `\\12\catcode `\$12\catcode
  `\&12\catcode `\#12\catcode `\^12\catcode `\_12\catcode `\%12\relax}%
\providecommand \@@startlink[1]{}%
\providecommand \@@endlink[0]{}%
\providecommand \url  [0]{\begingroup\@sanitize@url \@url }%
\providecommand \@url [1]{\endgroup\@href {#1}{\urlprefix }}%
\providecommand \urlprefix  [0]{URL }%
\providecommand \Eprint [0]{\href }%
\providecommand \doibase [0]{https://doi.org/}%
\providecommand \selectlanguage [0]{\@gobble}%
\providecommand \bibinfo  [0]{\@secondoftwo}%
\providecommand \bibfield  [0]{\@secondoftwo}%
\providecommand \translation [1]{[#1]}%
\providecommand \BibitemOpen [0]{}%
\providecommand \bibitemStop [0]{}%
\providecommand \bibitemNoStop [0]{.\EOS\space}%
\providecommand \EOS [0]{\spacefactor3000\relax}%
\providecommand \BibitemShut  [1]{\csname bibitem#1\endcsname}%
\let\auto@bib@innerbib\@empty
\bibitem [{\citenamefont {Buchenau}\ \emph {et~al.}(1991)\citenamefont
  {Buchenau}, \citenamefont {Galperin}, \citenamefont {Gurevich},\ and\
  \citenamefont {Schober}}]{soft_potential_model_1991}%
  \BibitemOpen
  \bibfield  {author} {\bibinfo {author} {\bibfnamefont {U.}~\bibnamefont
  {Buchenau}}, \bibinfo {author} {\bibfnamefont {Y.~M.}\ \bibnamefont
  {Galperin}}, \bibinfo {author} {\bibfnamefont {V.~L.}\ \bibnamefont
  {Gurevich}},\ and\ \bibinfo {author} {\bibfnamefont {H.~R.}\ \bibnamefont
  {Schober}},\ }\bibfield  {title} {\bibinfo {title} {Anharmonic potentials and
  vibrational localization in glasses},\ }\href
  {https://doi.org/10.1103/PhysRevB.43.5039} {\bibfield  {journal} {\bibinfo
  {journal} {Phys. Rev. B}\ }\textbf {\bibinfo {volume} {43}},\ \bibinfo
  {pages} {5039} (\bibinfo {year} {1991})}\BibitemShut {NoStop}%
\bibitem [{\citenamefont {Buchenau}\ \emph {et~al.}(1992)\citenamefont
  {Buchenau}, \citenamefont {Galperin}, \citenamefont {Gurevich}, \citenamefont
  {Parshin}, \citenamefont {Ramos},\ and\ \citenamefont
  {Schober}}]{Schober_prb_1992}%
  \BibitemOpen
  \bibfield  {author} {\bibinfo {author} {\bibfnamefont {U.}~\bibnamefont
  {Buchenau}}, \bibinfo {author} {\bibfnamefont {Y.~M.}\ \bibnamefont
  {Galperin}}, \bibinfo {author} {\bibfnamefont {V.~L.}\ \bibnamefont
  {Gurevich}}, \bibinfo {author} {\bibfnamefont {D.~A.}\ \bibnamefont
  {Parshin}}, \bibinfo {author} {\bibfnamefont {M.~A.}\ \bibnamefont {Ramos}},\
  and\ \bibinfo {author} {\bibfnamefont {H.~R.}\ \bibnamefont {Schober}},\
  }\bibfield  {title} {\bibinfo {title} {Interaction of soft modes and sound
  waves in glasses},\ }\href {https://doi.org/10.1103/PhysRevB.46.2798}
  {\bibfield  {journal} {\bibinfo  {journal} {Phys. Rev. B}\ }\textbf {\bibinfo
  {volume} {46}},\ \bibinfo {pages} {2798} (\bibinfo {year}
  {1992})}\BibitemShut {NoStop}%
\bibitem [{\citenamefont {Gurevich}\ \emph {et~al.}(2003)\citenamefont
  {Gurevich}, \citenamefont {Parshin},\ and\ \citenamefont
  {Schober}}]{Gurevich2003}%
  \BibitemOpen
  \bibfield  {author} {\bibinfo {author} {\bibfnamefont {V.~L.}\ \bibnamefont
  {Gurevich}}, \bibinfo {author} {\bibfnamefont {D.~A.}\ \bibnamefont
  {Parshin}},\ and\ \bibinfo {author} {\bibfnamefont {H.~R.}\ \bibnamefont
  {Schober}},\ }\bibfield  {title} {\bibinfo {title} {Anharmonicity,
  vibrational instability, and the boson peak in glasses},\ }\href
  {https://doi.org/10.1103/PhysRevB.67.094203} {\bibfield  {journal} {\bibinfo
  {journal} {Phys. Rev. B}\ }\textbf {\bibinfo {volume} {67}},\ \bibinfo
  {pages} {094203} (\bibinfo {year} {2003})}\BibitemShut {NoStop}%
\bibitem [{\citenamefont {Wyart}\ \emph {et~al.}(2005)\citenamefont {Wyart},
  \citenamefont {Silbert}, \citenamefont {Nagel},\ and\ \citenamefont
  {Witten}}]{matthieu_PRE_2005}%
  \BibitemOpen
  \bibfield  {author} {\bibinfo {author} {\bibfnamefont {M.}~\bibnamefont
  {Wyart}}, \bibinfo {author} {\bibfnamefont {L.~E.}\ \bibnamefont {Silbert}},
  \bibinfo {author} {\bibfnamefont {S.~R.}\ \bibnamefont {Nagel}},\ and\
  \bibinfo {author} {\bibfnamefont {T.~A.}\ \bibnamefont {Witten}},\ }\bibfield
   {title} {\bibinfo {title} {Effects of compression on the vibrational modes
  of marginally jammed solids},\ }\href
  {https://doi.org/10.1103/PhysRevE.72.051306} {\bibfield  {journal} {\bibinfo
  {journal} {Phys. Rev. E}\ }\textbf {\bibinfo {volume} {72}},\ \bibinfo
  {pages} {051306} (\bibinfo {year} {2005})}\BibitemShut {NoStop}%
\bibitem [{\citenamefont {Schirmacher}\ \emph {et~al.}(2007)\citenamefont
  {Schirmacher}, \citenamefont {Ruocco},\ and\ \citenamefont
  {Scopigno}}]{Schirmacher_prl_2007}%
  \BibitemOpen
  \bibfield  {author} {\bibinfo {author} {\bibfnamefont {W.}~\bibnamefont
  {Schirmacher}}, \bibinfo {author} {\bibfnamefont {G.}~\bibnamefont
  {Ruocco}},\ and\ \bibinfo {author} {\bibfnamefont {T.}~\bibnamefont
  {Scopigno}},\ }\bibfield  {title} {\bibinfo {title} {Acoustic attenuation in
  glasses and its relation with the boson peak},\ }\href
  {https://doi.org/10.1103/PhysRevLett.98.025501} {\bibfield  {journal}
  {\bibinfo  {journal} {Phys. Rev. Lett.}\ }\textbf {\bibinfo {volume} {98}},\
  \bibinfo {pages} {025501} (\bibinfo {year} {2007})}\BibitemShut {NoStop}%
\bibitem [{\citenamefont {DeGiuli}\ \emph {et~al.}(2014)\citenamefont
  {DeGiuli}, \citenamefont {Laversanne-Finot}, \citenamefont {During},
  \citenamefont {Lerner},\ and\ \citenamefont {Wyart}}]{eric_boson_peak_emt}%
  \BibitemOpen
  \bibfield  {author} {\bibinfo {author} {\bibfnamefont {E.}~\bibnamefont
  {DeGiuli}}, \bibinfo {author} {\bibfnamefont {A.}~\bibnamefont
  {Laversanne-Finot}}, \bibinfo {author} {\bibfnamefont {G.}~\bibnamefont
  {During}}, \bibinfo {author} {\bibfnamefont {E.}~\bibnamefont {Lerner}},\
  and\ \bibinfo {author} {\bibfnamefont {M.}~\bibnamefont {Wyart}},\ }\bibfield
   {title} {\bibinfo {title} {Effects of coordination and pressure on sound
  attenuation{,} boson peak and elasticity in amorphous solids},\ }\href
  {https://doi.org/10.1039/C4SM00561A} {\bibfield  {journal} {\bibinfo
  {journal} {Soft Matter}\ }\textbf {\bibinfo {volume} {10}},\ \bibinfo {pages}
  {5628} (\bibinfo {year} {2014})}\BibitemShut {NoStop}%
\bibitem [{\citenamefont {Franz}\ \emph {et~al.}(2015)\citenamefont {Franz},
  \citenamefont {Parisi}, \citenamefont {Urbani},\ and\ \citenamefont
  {Zamponi}}]{silvio}%
  \BibitemOpen
  \bibfield  {author} {\bibinfo {author} {\bibfnamefont {S.}~\bibnamefont
  {Franz}}, \bibinfo {author} {\bibfnamefont {G.}~\bibnamefont {Parisi}},
  \bibinfo {author} {\bibfnamefont {P.}~\bibnamefont {Urbani}},\ and\ \bibinfo
  {author} {\bibfnamefont {F.}~\bibnamefont {Zamponi}},\ }\bibfield  {title}
  {\bibinfo {title} {Universal spectrum of normal modes in low-temperature
  glasses},\ }\href {https://doi.org/10.1073/pnas.1511134112} {\bibfield
  {journal} {\bibinfo  {journal} {Proc. Natl. Acad. Sci. U.S.A.}\ }\textbf
  {\bibinfo {volume} {112}},\ \bibinfo {pages} {14539} (\bibinfo {year}
  {2015})}\BibitemShut {NoStop}%
\bibitem [{\citenamefont {Laird}\ and\ \citenamefont
  {Schober}(1991)}]{Schober_Laird_numerics_PRL}%
  \BibitemOpen
  \bibfield  {author} {\bibinfo {author} {\bibfnamefont {B.~B.}\ \bibnamefont
  {Laird}}\ and\ \bibinfo {author} {\bibfnamefont {H.~R.}\ \bibnamefont
  {Schober}},\ }\bibfield  {title} {\bibinfo {title} {Localized low-frequency
  vibrational modes in a simple model glass},\ }\href
  {https://doi.org/10.1103/PhysRevLett.66.636} {\bibfield  {journal} {\bibinfo
  {journal} {Phys. Rev. Lett.}\ }\textbf {\bibinfo {volume} {66}},\ \bibinfo
  {pages} {636} (\bibinfo {year} {1991})}\BibitemShut {NoStop}%
\bibitem [{\citenamefont {Schober}\ and\ \citenamefont
  {Oligschleger}(1996)}]{Schober_Oligschleger_numerics_PRB}%
  \BibitemOpen
  \bibfield  {author} {\bibinfo {author} {\bibfnamefont {H.~R.}\ \bibnamefont
  {Schober}}\ and\ \bibinfo {author} {\bibfnamefont {C.}~\bibnamefont
  {Oligschleger}},\ }\bibfield  {title} {\bibinfo {title} {Low-frequency
  vibrations in a model glass},\ }\href
  {https://doi.org/10.1103/PhysRevB.53.11469} {\bibfield  {journal} {\bibinfo
  {journal} {Phys. Rev. B}\ }\textbf {\bibinfo {volume} {53}},\ \bibinfo
  {pages} {11469} (\bibinfo {year} {1996})}\BibitemShut {NoStop}%
\bibitem [{\citenamefont {O'Hern}\ \emph {et~al.}(2003)\citenamefont {O'Hern},
  \citenamefont {Silbert}, \citenamefont {Liu},\ and\ \citenamefont
  {Nagel}}]{ohern2003}%
  \BibitemOpen
  \bibfield  {author} {\bibinfo {author} {\bibfnamefont {C.~S.}\ \bibnamefont
  {O'Hern}}, \bibinfo {author} {\bibfnamefont {L.~E.}\ \bibnamefont {Silbert}},
  \bibinfo {author} {\bibfnamefont {A.~J.}\ \bibnamefont {Liu}},\ and\ \bibinfo
  {author} {\bibfnamefont {S.~R.}\ \bibnamefont {Nagel}},\ }\bibfield  {title}
  {\bibinfo {title} {Jamming at zero temperature and zero applied stress: The
  epitome of disorder},\ }\href {https://doi.org/10.1103/PhysRevE.68.011306}
  {\bibfield  {journal} {\bibinfo  {journal} {Phys. Rev. E}\ }\textbf {\bibinfo
  {volume} {68}},\ \bibinfo {pages} {011306} (\bibinfo {year}
  {2003})}\BibitemShut {NoStop}%
\bibitem [{\citenamefont {Leonforte}\ \emph {et~al.}(2005)\citenamefont
  {Leonforte}, \citenamefont {Boissi\`ere}, \citenamefont {Tanguy},
  \citenamefont {Wittmer},\ and\ \citenamefont {Barrat}}]{barrat_3d}%
  \BibitemOpen
  \bibfield  {author} {\bibinfo {author} {\bibfnamefont {F.}~\bibnamefont
  {Leonforte}}, \bibinfo {author} {\bibfnamefont {R.}~\bibnamefont
  {Boissi\`ere}}, \bibinfo {author} {\bibfnamefont {A.}~\bibnamefont {Tanguy}},
  \bibinfo {author} {\bibfnamefont {J.~P.}\ \bibnamefont {Wittmer}},\ and\
  \bibinfo {author} {\bibfnamefont {J.-L.}\ \bibnamefont {Barrat}},\ }\bibfield
   {title} {\bibinfo {title} {Continuum limit of amorphous elastic bodies. iii.
  three-dimensional systems},\ }\href
  {https://doi.org/10.1103/PhysRevB.72.224206} {\bibfield  {journal} {\bibinfo
  {journal} {Phys. Rev. B}\ }\textbf {\bibinfo {volume} {72}},\ \bibinfo
  {pages} {224206} (\bibinfo {year} {2005})}\BibitemShut {NoStop}%
\bibitem [{\citenamefont {Xu}\ \emph {et~al.}(2010)\citenamefont {Xu},
  \citenamefont {Vitelli}, \citenamefont {Liu},\ and\ \citenamefont
  {Nagel}}]{vincenzo_epl_2010}%
  \BibitemOpen
  \bibfield  {author} {\bibinfo {author} {\bibfnamefont {N.}~\bibnamefont
  {Xu}}, \bibinfo {author} {\bibfnamefont {V.}~\bibnamefont {Vitelli}},
  \bibinfo {author} {\bibfnamefont {A.~J.}\ \bibnamefont {Liu}},\ and\ \bibinfo
  {author} {\bibfnamefont {S.~R.}\ \bibnamefont {Nagel}},\ }\bibfield  {title}
  {\bibinfo {title} {Anharmonic and quasi-localized vibrations in jammed solids
  {--} modes for mechanical failure},\ }\href
  {http://stacks.iop.org/0295-5075/90/i=5/a=56001} {\bibfield  {journal}
  {\bibinfo  {journal} {Europhys. Lett.}\ }\textbf {\bibinfo {volume} {90}},\
  \bibinfo {pages} {56001} (\bibinfo {year} {2010})}\BibitemShut {NoStop}%
\bibitem [{\citenamefont {Lerner}\ \emph {et~al.}(2016)\citenamefont {Lerner},
  \citenamefont {D\"uring},\ and\ \citenamefont {Bouchbinder}}]{modes_prl}%
  \BibitemOpen
  \bibfield  {author} {\bibinfo {author} {\bibfnamefont {E.}~\bibnamefont
  {Lerner}}, \bibinfo {author} {\bibfnamefont {G.}~\bibnamefont {D\"uring}},\
  and\ \bibinfo {author} {\bibfnamefont {E.}~\bibnamefont {Bouchbinder}},\
  }\bibfield  {title} {\bibinfo {title} {Statistics and properties of
  low-frequency vibrational modes in structural glasses},\ }\href
  {https://doi.org/10.1103/PhysRevLett.117.035501} {\bibfield  {journal}
  {\bibinfo  {journal} {Phys. Rev. Lett.}\ }\textbf {\bibinfo {volume} {117}},\
  \bibinfo {pages} {035501} (\bibinfo {year} {2016})}\BibitemShut {NoStop}%
\bibitem [{\citenamefont {Kapteijns}\ \emph {et~al.}(2018)\citenamefont
  {Kapteijns}, \citenamefont {Bouchbinder},\ and\ \citenamefont
  {Lerner}}]{modes_prl_2018}%
  \BibitemOpen
  \bibfield  {author} {\bibinfo {author} {\bibfnamefont {G.}~\bibnamefont
  {Kapteijns}}, \bibinfo {author} {\bibfnamefont {E.}~\bibnamefont
  {Bouchbinder}},\ and\ \bibinfo {author} {\bibfnamefont {E.}~\bibnamefont
  {Lerner}},\ }\bibfield  {title} {\bibinfo {title} {Universal nonphononic
  density of states in 2d, 3d, and 4d glasses},\ }\href
  {https://doi.org/10.1103/PhysRevLett.121.055501} {\bibfield  {journal}
  {\bibinfo  {journal} {Phys. Rev. Lett.}\ }\textbf {\bibinfo {volume} {121}},\
  \bibinfo {pages} {055501} (\bibinfo {year} {2018})}\BibitemShut {NoStop}%
\bibitem [{\citenamefont {Mizuno}\ \emph {et~al.}(2017)\citenamefont {Mizuno},
  \citenamefont {Shiba},\ and\ \citenamefont {Ikeda}}]{ikeda_pnas}%
  \BibitemOpen
  \bibfield  {author} {\bibinfo {author} {\bibfnamefont {H.}~\bibnamefont
  {Mizuno}}, \bibinfo {author} {\bibfnamefont {H.}~\bibnamefont {Shiba}},\ and\
  \bibinfo {author} {\bibfnamefont {A.}~\bibnamefont {Ikeda}},\ }\bibfield
  {title} {\bibinfo {title} {Continuum limit of the vibrational properties of
  amorphous solids},\ }\href {https://doi.org/10.1073/pnas.1709015114}
  {\bibfield  {journal} {\bibinfo  {journal} {Proc. Natl. Acad. Sci. U.S.A.}\
  }\textbf {\bibinfo {volume} {114}},\ \bibinfo {pages} {E9767} (\bibinfo
  {year} {2017})}\BibitemShut {NoStop}%
\bibitem [{\citenamefont {Monaco}\ \emph {et~al.}(2006)\citenamefont {Monaco},
  \citenamefont {Chumakov}, \citenamefont {Monaco}, \citenamefont {Crichton},
  \citenamefont {Meyer}, \citenamefont {Comez}, \citenamefont {Fioretto},
  \citenamefont {Korecki},\ and\ \citenamefont
  {R\"uffer}}]{Effect_of_Densification_prl_2006}%
  \BibitemOpen
  \bibfield  {author} {\bibinfo {author} {\bibfnamefont {A.}~\bibnamefont
  {Monaco}}, \bibinfo {author} {\bibfnamefont {A.~I.}\ \bibnamefont
  {Chumakov}}, \bibinfo {author} {\bibfnamefont {G.}~\bibnamefont {Monaco}},
  \bibinfo {author} {\bibfnamefont {W.~A.}\ \bibnamefont {Crichton}}, \bibinfo
  {author} {\bibfnamefont {A.}~\bibnamefont {Meyer}}, \bibinfo {author}
  {\bibfnamefont {L.}~\bibnamefont {Comez}}, \bibinfo {author} {\bibfnamefont
  {D.}~\bibnamefont {Fioretto}}, \bibinfo {author} {\bibfnamefont
  {J.}~\bibnamefont {Korecki}},\ and\ \bibinfo {author} {\bibfnamefont
  {R.}~\bibnamefont {R\"uffer}},\ }\bibfield  {title} {\bibinfo {title} {Effect
  of densification on the density of vibrational states of glasses},\ }\href
  {https://doi.org/10.1103/PhysRevLett.97.135501} {\bibfield  {journal}
  {\bibinfo  {journal} {Phys. Rev. Lett.}\ }\textbf {\bibinfo {volume} {97}},\
  \bibinfo {pages} {135501} (\bibinfo {year} {2006})}\BibitemShut {NoStop}%
\bibitem [{\citenamefont {Baldi}\ \emph {et~al.}(2011)\citenamefont {Baldi},
  \citenamefont {Giordano},\ and\ \citenamefont
  {Monaco}}]{experiments_300K_vSiO2}%
  \BibitemOpen
  \bibfield  {author} {\bibinfo {author} {\bibfnamefont {G.}~\bibnamefont
  {Baldi}}, \bibinfo {author} {\bibfnamefont {V.~M.}\ \bibnamefont
  {Giordano}},\ and\ \bibinfo {author} {\bibfnamefont {G.}~\bibnamefont
  {Monaco}},\ }\bibfield  {title} {\bibinfo {title} {Elastic anomalies at
  terahertz frequencies and excess density of vibrational states in silica
  glass},\ }\href {https://doi.org/10.1103/PhysRevB.83.174203} {\bibfield
  {journal} {\bibinfo  {journal} {Phys. Rev. B}\ }\textbf {\bibinfo {volume}
  {83}},\ \bibinfo {pages} {174203} (\bibinfo {year} {2011})}\BibitemShut
  {NoStop}%
\bibitem [{\citenamefont {Baldi}\ \emph {et~al.}(2010)\citenamefont {Baldi},
  \citenamefont {Giordano}, \citenamefont {Monaco},\ and\ \citenamefont
  {Ruta}}]{experiments_1620K_vSiO2}%
  \BibitemOpen
  \bibfield  {author} {\bibinfo {author} {\bibfnamefont {G.}~\bibnamefont
  {Baldi}}, \bibinfo {author} {\bibfnamefont {V.~M.}\ \bibnamefont {Giordano}},
  \bibinfo {author} {\bibfnamefont {G.}~\bibnamefont {Monaco}},\ and\ \bibinfo
  {author} {\bibfnamefont {B.}~\bibnamefont {Ruta}},\ }\bibfield  {title}
  {\bibinfo {title} {Sound attenuation at terahertz frequencies and the boson
  peak of vitreous silica},\ }\href
  {https://doi.org/10.1103/PhysRevLett.104.195501} {\bibfield  {journal}
  {\bibinfo  {journal} {Phys. Rev. Lett.}\ }\textbf {\bibinfo {volume} {104}},\
  \bibinfo {pages} {195501} (\bibinfo {year} {2010})}\BibitemShut {NoStop}%
\bibitem [{\citenamefont {Chumakov}\ \emph {et~al.}(2011)\citenamefont
  {Chumakov}, \citenamefont {Monaco}, \citenamefont {Monaco}, \citenamefont
  {Crichton}, \citenamefont {Bosak}, \citenamefont {R\"uffer}, \citenamefont
  {Meyer}, \citenamefont {Kargl}, \citenamefont {Comez}, \citenamefont
  {Fioretto}, \citenamefont {Giefers}, \citenamefont {Roitsch}, \citenamefont
  {Wortmann}, \citenamefont {Manghnani}, \citenamefont {Hushur}, \citenamefont
  {Williams}, \citenamefont {Balogh}, \citenamefont
  {Parli\ifmmode~\acute{n}\else \'{n}\fi{}ski}, \citenamefont {Jochym},\ and\
  \citenamefont {Piekarz}}]{Monaco_prl_2011}%
  \BibitemOpen
  \bibfield  {author} {\bibinfo {author} {\bibfnamefont {A.~I.}\ \bibnamefont
  {Chumakov}}, \bibinfo {author} {\bibfnamefont {G.}~\bibnamefont {Monaco}},
  \bibinfo {author} {\bibfnamefont {A.}~\bibnamefont {Monaco}}, \bibinfo
  {author} {\bibfnamefont {W.~A.}\ \bibnamefont {Crichton}}, \bibinfo {author}
  {\bibfnamefont {A.}~\bibnamefont {Bosak}}, \bibinfo {author} {\bibfnamefont
  {R.}~\bibnamefont {R\"uffer}}, \bibinfo {author} {\bibfnamefont
  {A.}~\bibnamefont {Meyer}}, \bibinfo {author} {\bibfnamefont
  {F.}~\bibnamefont {Kargl}}, \bibinfo {author} {\bibfnamefont
  {L.}~\bibnamefont {Comez}}, \bibinfo {author} {\bibfnamefont
  {D.}~\bibnamefont {Fioretto}}, \bibinfo {author} {\bibfnamefont
  {H.}~\bibnamefont {Giefers}}, \bibinfo {author} {\bibfnamefont
  {S.}~\bibnamefont {Roitsch}}, \bibinfo {author} {\bibfnamefont
  {G.}~\bibnamefont {Wortmann}}, \bibinfo {author} {\bibfnamefont {M.~H.}\
  \bibnamefont {Manghnani}}, \bibinfo {author} {\bibfnamefont {A.}~\bibnamefont
  {Hushur}}, \bibinfo {author} {\bibfnamefont {Q.}~\bibnamefont {Williams}},
  \bibinfo {author} {\bibfnamefont {J.}~\bibnamefont {Balogh}}, \bibinfo
  {author} {\bibfnamefont {K.}~\bibnamefont {Parli\ifmmode~\acute{n}\else
  \'{n}\fi{}ski}}, \bibinfo {author} {\bibfnamefont {P.}~\bibnamefont
  {Jochym}},\ and\ \bibinfo {author} {\bibfnamefont {P.}~\bibnamefont
  {Piekarz}},\ }\bibfield  {title} {\bibinfo {title} {Equivalence of the boson
  peak in glasses to the transverse acoustic van hove singularity in
  crystals},\ }\href {https://doi.org/10.1103/PhysRevLett.106.225501}
  {\bibfield  {journal} {\bibinfo  {journal} {Phys. Rev. Lett.}\ }\textbf
  {\bibinfo {volume} {106}},\ \bibinfo {pages} {225501} (\bibinfo {year}
  {2011})}\BibitemShut {NoStop}%
\bibitem [{\citenamefont {Deng}\ \emph {et~al.}(1989)\citenamefont {Deng},
  \citenamefont {Argon},\ and\ \citenamefont {Yip}}]{argon_simulations}%
  \BibitemOpen
  \bibfield  {author} {\bibinfo {author} {\bibfnamefont {D.}~\bibnamefont
  {Deng}}, \bibinfo {author} {\bibfnamefont {A.~S.}\ \bibnamefont {Argon}},\
  and\ \bibinfo {author} {\bibfnamefont {S.}~\bibnamefont {Yip}},\ }\bibfield
  {title} {\bibinfo {title} {Simulation of plastic deformation in a
  two-dimensional atomic glass by molecular dynamics iv},\ }\href
  {https://doi.org/10.1098/rsta.1989.0092} {\bibfield  {journal} {\bibinfo
  {journal} {Philos. Trans. R. Soc. A}\ }\textbf {\bibinfo {volume} {329}},\
  \bibinfo {pages} {613} (\bibinfo {year} {1989})}\BibitemShut {NoStop}%
\bibitem [{\citenamefont {H\'ebraud}\ \emph {et~al.}(1997)\citenamefont
  {H\'ebraud}, \citenamefont {Lequeux}, \citenamefont {Munch},\ and\
  \citenamefont {Pine}}]{pine_emulsions_stzs}%
  \BibitemOpen
  \bibfield  {author} {\bibinfo {author} {\bibfnamefont {P.}~\bibnamefont
  {H\'ebraud}}, \bibinfo {author} {\bibfnamefont {F.}~\bibnamefont {Lequeux}},
  \bibinfo {author} {\bibfnamefont {J.~P.}\ \bibnamefont {Munch}},\ and\
  \bibinfo {author} {\bibfnamefont {D.~J.}\ \bibnamefont {Pine}},\ }\bibfield
  {title} {\bibinfo {title} {Yielding and rearrangements in disordered
  emulsions},\ }\href {https://doi.org/10.1103/PhysRevLett.78.4657} {\bibfield
  {journal} {\bibinfo  {journal} {Phys. Rev. Lett.}\ }\textbf {\bibinfo
  {volume} {78}},\ \bibinfo {pages} {4657} (\bibinfo {year}
  {1997})}\BibitemShut {NoStop}%
\bibitem [{\citenamefont {Varnik}\ \emph {et~al.}(2004)\citenamefont {Varnik},
  \citenamefont {Bocquet},\ and\ \citenamefont
  {Barrat}}]{barrat_yielding_jcp_2004}%
  \BibitemOpen
  \bibfield  {author} {\bibinfo {author} {\bibfnamefont {F.}~\bibnamefont
  {Varnik}}, \bibinfo {author} {\bibfnamefont {L.}~\bibnamefont {Bocquet}},\
  and\ \bibinfo {author} {\bibfnamefont {J.-L.}\ \bibnamefont {Barrat}},\
  }\bibfield  {title} {\bibinfo {title} {A study of the static yield stress in
  a binary lennard-jones glass},\ }\href {https://doi.org/10.1063/1.1636451}
  {\bibfield  {journal} {\bibinfo  {journal} {J. Chem. Phys.}\ }\textbf
  {\bibinfo {volume} {120}},\ \bibinfo {pages} {2788} (\bibinfo {year}
  {2004})}\BibitemShut {NoStop}%
\bibitem [{\citenamefont {Schall}\ \emph {et~al.}(2007)\citenamefont {Schall},
  \citenamefont {Weitz},\ and\ \citenamefont {Spaepen}}]{Schall1895}%
  \BibitemOpen
  \bibfield  {author} {\bibinfo {author} {\bibfnamefont {P.}~\bibnamefont
  {Schall}}, \bibinfo {author} {\bibfnamefont {D.~A.}\ \bibnamefont {Weitz}},\
  and\ \bibinfo {author} {\bibfnamefont {F.}~\bibnamefont {Spaepen}},\
  }\bibfield  {title} {\bibinfo {title} {Structural rearrangements that govern
  flow in colloidal glasses},\ }\href {https://doi.org/10.1126/science.1149308}
  {\bibfield  {journal} {\bibinfo  {journal} {Science}\ }\textbf {\bibinfo
  {volume} {318}},\ \bibinfo {pages} {1895} (\bibinfo {year}
  {2007})}\BibitemShut {NoStop}%
\bibitem [{\citenamefont {Ozawa}\ \emph {et~al.}(2018)\citenamefont {Ozawa},
  \citenamefont {Berthier}, \citenamefont {Biroli}, \citenamefont {Rosso},\
  and\ \citenamefont {Tarjus}}]{Ozawa6656}%
  \BibitemOpen
  \bibfield  {author} {\bibinfo {author} {\bibfnamefont {M.}~\bibnamefont
  {Ozawa}}, \bibinfo {author} {\bibfnamefont {L.}~\bibnamefont {Berthier}},
  \bibinfo {author} {\bibfnamefont {G.}~\bibnamefont {Biroli}}, \bibinfo
  {author} {\bibfnamefont {A.}~\bibnamefont {Rosso}},\ and\ \bibinfo {author}
  {\bibfnamefont {G.}~\bibnamefont {Tarjus}},\ }\bibfield  {title} {\bibinfo
  {title} {Random critical point separates brittle and ductile yielding
  transitions in amorphous materials},\ }\href
  {https://doi.org/10.1073/pnas.1806156115} {\bibfield  {journal} {\bibinfo
  {journal} {Proc. Natl. Acad. Sci. U.S.A.}\ }\textbf {\bibinfo {volume}
  {115}},\ \bibinfo {pages} {6656} (\bibinfo {year} {2018})}\BibitemShut
  {NoStop}%
\bibitem [{\citenamefont {Alix-Williams}\ and\ \citenamefont
  {Falk}(2018)}]{falk_shear_banding_pre_2018}%
  \BibitemOpen
  \bibfield  {author} {\bibinfo {author} {\bibfnamefont {D.~D.}\ \bibnamefont
  {Alix-Williams}}\ and\ \bibinfo {author} {\bibfnamefont {M.~L.}\ \bibnamefont
  {Falk}},\ }\bibfield  {title} {\bibinfo {title} {Shear band broadening in
  simulated glasses},\ }\href {https://doi.org/10.1103/PhysRevE.98.053002}
  {\bibfield  {journal} {\bibinfo  {journal} {Phys. Rev. E}\ }\textbf {\bibinfo
  {volume} {98}},\ \bibinfo {pages} {053002} (\bibinfo {year}
  {2018})}\BibitemShut {NoStop}%
\bibitem [{\citenamefont {Popovi\ifmmode~\acute{c}\else \'{c}\fi{}}\ \emph
  {et~al.}(2018)\citenamefont {Popovi\ifmmode~\acute{c}\else \'{c}\fi{}},
  \citenamefont {de~Geus},\ and\ \citenamefont {Wyart}}]{MW_yielding_2018_pre}%
  \BibitemOpen
  \bibfield  {author} {\bibinfo {author} {\bibfnamefont {M.}~\bibnamefont
  {Popovi\ifmmode~\acute{c}\else \'{c}\fi{}}}, \bibinfo {author} {\bibfnamefont
  {T.~W.~J.}\ \bibnamefont {de~Geus}},\ and\ \bibinfo {author} {\bibfnamefont
  {M.}~\bibnamefont {Wyart}},\ }\bibfield  {title} {\bibinfo {title}
  {Elastoplastic description of sudden failure in athermal amorphous materials
  during quasistatic loading},\ }\href
  {https://doi.org/10.1103/PhysRevE.98.040901} {\bibfield  {journal} {\bibinfo
  {journal} {Phys. Rev. E}\ }\textbf {\bibinfo {volume} {98}},\ \bibinfo
  {pages} {040901} (\bibinfo {year} {2018})}\BibitemShut {NoStop}%
\bibitem [{\citenamefont {Barbot}\ \emph {et~al.}(2020)\citenamefont {Barbot},
  \citenamefont {Lerbinger}, \citenamefont {Lema\^{\i}tre}, \citenamefont
  {Vandembroucq},\ and\ \citenamefont {Patinet}}]{Sylvain_pre_rejuvenation}%
  \BibitemOpen
  \bibfield  {author} {\bibinfo {author} {\bibfnamefont {A.}~\bibnamefont
  {Barbot}}, \bibinfo {author} {\bibfnamefont {M.}~\bibnamefont {Lerbinger}},
  \bibinfo {author} {\bibfnamefont {A.}~\bibnamefont {Lema\^{\i}tre}}, \bibinfo
  {author} {\bibfnamefont {D.}~\bibnamefont {Vandembroucq}},\ and\ \bibinfo
  {author} {\bibfnamefont {S.}~\bibnamefont {Patinet}},\ }\bibfield  {title}
  {\bibinfo {title} {Rejuvenation and shear banding in model amorphous
  solids},\ }\href {https://doi.org/10.1103/PhysRevE.101.033001} {\bibfield
  {journal} {\bibinfo  {journal} {Phys. Rev. E}\ }\textbf {\bibinfo {volume}
  {101}},\ \bibinfo {pages} {033001} (\bibinfo {year} {2020})}\BibitemShut
  {NoStop}%
\bibitem [{\citenamefont {Moriel}\ \emph {et~al.}(2019)\citenamefont {Moriel},
  \citenamefont {Kapteijns}, \citenamefont {Rainone}, \citenamefont {Zylberg},
  \citenamefont {Lerner},\ and\ \citenamefont {Bouchbinder}}]{scattering_jcp}%
  \BibitemOpen
  \bibfield  {author} {\bibinfo {author} {\bibfnamefont {A.}~\bibnamefont
  {Moriel}}, \bibinfo {author} {\bibfnamefont {G.}~\bibnamefont {Kapteijns}},
  \bibinfo {author} {\bibfnamefont {C.}~\bibnamefont {Rainone}}, \bibinfo
  {author} {\bibfnamefont {J.}~\bibnamefont {Zylberg}}, \bibinfo {author}
  {\bibfnamefont {E.}~\bibnamefont {Lerner}},\ and\ \bibinfo {author}
  {\bibfnamefont {E.}~\bibnamefont {Bouchbinder}},\ }\bibfield  {title}
  {\bibinfo {title} {Wave attenuation in glasses: Rayleigh and
  generalized-rayleigh scattering scaling},\ }\href
  {https://doi.org/10.1063/1.5111192} {\bibfield  {journal} {\bibinfo
  {journal} {J. Chem. Phys.}\ }\textbf {\bibinfo {volume} {151}},\ \bibinfo
  {pages} {104503} (\bibinfo {year} {2019})}\BibitemShut {NoStop}%
\bibitem [{\citenamefont {Schober}\ \emph {et~al.}(1993)\citenamefont
  {Schober}, \citenamefont {Oligschleger},\ and\ \citenamefont
  {Laird}}]{schober1993_numerics}%
  \BibitemOpen
  \bibfield  {author} {\bibinfo {author} {\bibfnamefont {H.}~\bibnamefont
  {Schober}}, \bibinfo {author} {\bibfnamefont {C.}~\bibnamefont
  {Oligschleger}},\ and\ \bibinfo {author} {\bibfnamefont {B.}~\bibnamefont
  {Laird}},\ }\bibfield  {title} {\bibinfo {title} {Low-frequency vibrations
  and relaxations in glasses},\ }\href
  {https://doi.org/https://doi.org/10.1016/0022-3093(93)90106-8} {\bibfield
  {journal} {\bibinfo  {journal} {J. Non-Cryst. Solids}\ }\textbf {\bibinfo
  {volume} {156-158}},\ \bibinfo {pages} {965 } (\bibinfo {year}
  {1993})}\BibitemShut {NoStop}%
\bibitem [{\citenamefont {Oligschleger}\ and\ \citenamefont
  {Schober}(1999)}]{Schober_correlate_modes_dynamics}%
  \BibitemOpen
  \bibfield  {author} {\bibinfo {author} {\bibfnamefont {C.}~\bibnamefont
  {Oligschleger}}\ and\ \bibinfo {author} {\bibfnamefont {H.~R.}\ \bibnamefont
  {Schober}},\ }\bibfield  {title} {\bibinfo {title} {Collective jumps in a
  soft-sphere glass},\ }\href {https://doi.org/10.1103/PhysRevB.59.811}
  {\bibfield  {journal} {\bibinfo  {journal} {Phys. Rev. B}\ }\textbf {\bibinfo
  {volume} {59}},\ \bibinfo {pages} {811} (\bibinfo {year} {1999})}\BibitemShut
  {NoStop}%
\bibitem [{\citenamefont {Widmer-Cooper}\ \emph {et~al.}(2008)\citenamefont
  {Widmer-Cooper}, \citenamefont {Perry}, \citenamefont {Harrowell},\ and\
  \citenamefont {Reichman}}]{widmer2008irreversible}%
  \BibitemOpen
  \bibfield  {author} {\bibinfo {author} {\bibfnamefont {A.}~\bibnamefont
  {Widmer-Cooper}}, \bibinfo {author} {\bibfnamefont {H.}~\bibnamefont
  {Perry}}, \bibinfo {author} {\bibfnamefont {P.}~\bibnamefont {Harrowell}},\
  and\ \bibinfo {author} {\bibfnamefont {D.~R.}\ \bibnamefont {Reichman}},\
  }\bibfield  {title} {\bibinfo {title} {Irreversible reorganization in a
  supercooled liquid originates from localized soft modes},\ }\href
  {https://doi.org/10.1038/nphys1025} {\bibfield  {journal} {\bibinfo
  {journal} {Nature Phys.}\ }\textbf {\bibinfo {volume} {4}},\ \bibinfo {pages}
  {711} (\bibinfo {year} {2008})}\BibitemShut {NoStop}%
\bibitem [{\citenamefont {Anderson}\ \emph {et~al.}(1972)\citenamefont
  {Anderson}, \citenamefont {Halperin},\ and\ \citenamefont
  {Varma}}]{Anderson}%
  \BibitemOpen
  \bibfield  {author} {\bibinfo {author} {\bibfnamefont {P.~W.}\ \bibnamefont
  {Anderson}}, \bibinfo {author} {\bibfnamefont {B.~I.}\ \bibnamefont
  {Halperin}},\ and\ \bibinfo {author} {\bibfnamefont {C.~M.}\ \bibnamefont
  {Varma}},\ }\bibfield  {title} {\bibinfo {title} {Anomalous low-temperature
  thermal properties of glasses and spin glasses},\ }\href
  {https://doi.org/10.1080/14786437208229210} {\bibfield  {journal} {\bibinfo
  {journal} {Philos. Mag.}\ }\textbf {\bibinfo {volume} {25}},\ \bibinfo
  {pages} {1} (\bibinfo {year} {1972})}\BibitemShut {NoStop}%
\bibitem [{\citenamefont {Phillips}(1972)}]{Phillips}%
  \BibitemOpen
  \bibfield  {author} {\bibinfo {author} {\bibfnamefont {W.}~\bibnamefont
  {Phillips}},\ }\bibfield  {title} {\bibinfo {title} {Tunneling states in
  amorphous solids},\ }\href {https://doi.org/10.1007/BF00660072} {\bibfield
  {journal} {\bibinfo  {journal} {J. Low Temp. Phys.}\ }\textbf {\bibinfo
  {volume} {7}},\ \bibinfo {pages} {351} (\bibinfo {year} {1972})}\BibitemShut
  {NoStop}%
\bibitem [{\citenamefont {Khomenko}\ \emph {et~al.}(2019)\citenamefont
  {Khomenko}, \citenamefont {Scalliet}, \citenamefont {Berthier}, \citenamefont
  {Reichman},\ and\ \citenamefont {Zamponi}}]{khomenko2019depletion}%
  \BibitemOpen
  \bibfield  {author} {\bibinfo {author} {\bibfnamefont {D.}~\bibnamefont
  {Khomenko}}, \bibinfo {author} {\bibfnamefont {C.}~\bibnamefont {Scalliet}},
  \bibinfo {author} {\bibfnamefont {L.}~\bibnamefont {Berthier}}, \bibinfo
  {author} {\bibfnamefont {D.~R.}\ \bibnamefont {Reichman}},\ and\ \bibinfo
  {author} {\bibfnamefont {F.}~\bibnamefont {Zamponi}},\ }\bibfield  {title}
  {\bibinfo {title} {Depletion of two-level systems in ultrastable
  computer-generated glasses},\ }\href {https://arxiv.org/abs/1910.11168}
  {\bibfield  {journal} {\bibinfo  {journal} {arXiv preprint arXiv:1910.11168}\
  } (\bibinfo {year} {2019})}\BibitemShut {NoStop}%
\bibitem [{\citenamefont {Shimada}\ \emph {et~al.}(2018)\citenamefont
  {Shimada}, \citenamefont {Mizuno}, \citenamefont {Wyart},\ and\ \citenamefont
  {Ikeda}}]{atsushi_core_size_pre}%
  \BibitemOpen
  \bibfield  {author} {\bibinfo {author} {\bibfnamefont {M.}~\bibnamefont
  {Shimada}}, \bibinfo {author} {\bibfnamefont {H.}~\bibnamefont {Mizuno}},
  \bibinfo {author} {\bibfnamefont {M.}~\bibnamefont {Wyart}},\ and\ \bibinfo
  {author} {\bibfnamefont {A.}~\bibnamefont {Ikeda}},\ }\bibfield  {title}
  {\bibinfo {title} {Spatial structure of quasilocalized vibrations in nearly
  jammed amorphous solids},\ }\href
  {https://doi.org/10.1103/PhysRevE.98.060901} {\bibfield  {journal} {\bibinfo
  {journal} {Phys. Rev. E}\ }\textbf {\bibinfo {volume} {98}},\ \bibinfo
  {pages} {060901} (\bibinfo {year} {2018})}\BibitemShut {NoStop}%
\bibitem [{\citenamefont {Rainone}\ \emph {et~al.}(2020)\citenamefont
  {Rainone}, \citenamefont {Bouchbinder},\ and\ \citenamefont
  {Lerner}}]{pinching_2019}%
  \BibitemOpen
  \bibfield  {author} {\bibinfo {author} {\bibfnamefont {C.}~\bibnamefont
  {Rainone}}, \bibinfo {author} {\bibfnamefont {E.}~\bibnamefont
  {Bouchbinder}},\ and\ \bibinfo {author} {\bibfnamefont {E.}~\bibnamefont
  {Lerner}},\ }\bibfield  {title} {\bibinfo {title} {Pinching a glass reveals
  key properties of its soft spots},\ }\href
  {https://doi.org/10.1073/pnas.1919958117} {\bibfield  {journal} {\bibinfo
  {journal} {Proc. Natl. Acad. Sci. U.S.A.}\ } (\bibinfo {year}
  {2020})}\BibitemShut {NoStop}%
\bibitem [{\citenamefont {Lerner}\ and\ \citenamefont
  {Bouchbinder}(2017)}]{protocol_prerc}%
  \BibitemOpen
  \bibfield  {author} {\bibinfo {author} {\bibfnamefont {E.}~\bibnamefont
  {Lerner}}\ and\ \bibinfo {author} {\bibfnamefont {E.}~\bibnamefont
  {Bouchbinder}},\ }\bibfield  {title} {\bibinfo {title} {Effect of
  instantaneous and continuous quenches on the density of vibrational modes in
  model glasses},\ }\href {https://doi.org/10.1103/PhysRevE.96.020104}
  {\bibfield  {journal} {\bibinfo  {journal} {Phys. Rev. E}\ }\textbf {\bibinfo
  {volume} {96}},\ \bibinfo {pages} {020104} (\bibinfo {year}
  {2017})}\BibitemShut {NoStop}%
\bibitem [{\citenamefont {Lerner}\ and\ \citenamefont
  {Bouchbinder}(2018)}]{cge_paper}%
  \BibitemOpen
  \bibfield  {author} {\bibinfo {author} {\bibfnamefont {E.}~\bibnamefont
  {Lerner}}\ and\ \bibinfo {author} {\bibfnamefont {E.}~\bibnamefont
  {Bouchbinder}},\ }\bibfield  {title} {\bibinfo {title} {A characteristic
  energy scale in glasses},\ }\href {https://doi.org/10.1063/1.5024776}
  {\bibfield  {journal} {\bibinfo  {journal} {J. Chem. Phys.}\ }\textbf
  {\bibinfo {volume} {148}},\ \bibinfo {pages} {214502} (\bibinfo {year}
  {2018})}\BibitemShut {NoStop}%
\bibitem [{\citenamefont {Wang}\ \emph {et~al.}(2019)\citenamefont {Wang},
  \citenamefont {Ninarello}, \citenamefont {Guan}, \citenamefont {Berthier},
  \citenamefont {Szamel},\ and\ \citenamefont {Flenner}}]{LB_modes_2019}%
  \BibitemOpen
  \bibfield  {author} {\bibinfo {author} {\bibfnamefont {L.}~\bibnamefont
  {Wang}}, \bibinfo {author} {\bibfnamefont {A.}~\bibnamefont {Ninarello}},
  \bibinfo {author} {\bibfnamefont {P.}~\bibnamefont {Guan}}, \bibinfo {author}
  {\bibfnamefont {L.}~\bibnamefont {Berthier}}, \bibinfo {author}
  {\bibfnamefont {G.}~\bibnamefont {Szamel}},\ and\ \bibinfo {author}
  {\bibfnamefont {E.}~\bibnamefont {Flenner}},\ }\bibfield  {title} {\bibinfo
  {title} {Low-frequency vibrational modes of stable glasses},\ }\href
  {https://doi.org/10.1038/s41467-018-07978-1} {\bibfield  {journal} {\bibinfo
  {journal} {Nat. Commun.}\ }\textbf {\bibinfo {volume} {10}},\ \bibinfo
  {pages} {26} (\bibinfo {year} {2019})}\BibitemShut {NoStop}%
\bibitem [{\citenamefont {Lehoucq}\ \emph {et~al.}(1998)\citenamefont
  {Lehoucq}, \citenamefont {Sorensen},\ and\ \citenamefont {Yang}}]{arpack}%
  \BibitemOpen
  \bibfield  {author} {\bibinfo {author} {\bibfnamefont {R.~B.}\ \bibnamefont
  {Lehoucq}}, \bibinfo {author} {\bibfnamefont {D.~C.}\ \bibnamefont
  {Sorensen}},\ and\ \bibinfo {author} {\bibfnamefont {C.}~\bibnamefont
  {Yang}},\ }\href {https://doi.org/10.1137/1.9780898719628} {\emph {\bibinfo
  {title} {ARPACK Users' Guide}}}\ (\bibinfo  {publisher} {Society for
  Industrial and Applied Mathematics, Philadelphia},\ \bibinfo {year}
  {1998})\BibitemShut {NoStop}%
\bibitem [{\citenamefont {Bouchbinder}\ and\ \citenamefont
  {Lerner}(2018)}]{phonon_widths}%
  \BibitemOpen
  \bibfield  {author} {\bibinfo {author} {\bibfnamefont {E.}~\bibnamefont
  {Bouchbinder}}\ and\ \bibinfo {author} {\bibfnamefont {E.}~\bibnamefont
  {Lerner}},\ }\bibfield  {title} {\bibinfo {title} {Universal disorder-induced
  broadening of phonon bands: from disordered lattices to glasses},\ }\href
  {http://stacks.iop.org/1367-2630/20/i=7/a=073022} {\bibfield  {journal}
  {\bibinfo  {journal} {New J. Phys.}\ }\textbf {\bibinfo {volume} {20}},\
  \bibinfo {pages} {073022} (\bibinfo {year} {2018})}\BibitemShut {NoStop}%
\bibitem [{\citenamefont {Gartner}\ and\ \citenamefont
  {Lerner}(2016)}]{SciPost2016}%
  \BibitemOpen
  \bibfield  {author} {\bibinfo {author} {\bibfnamefont {L.}~\bibnamefont
  {Gartner}}\ and\ \bibinfo {author} {\bibfnamefont {E.}~\bibnamefont
  {Lerner}},\ }\bibfield  {title} {\bibinfo {title} {{Nonlinear modes
  disentangle glassy and Goldstone modes in structural glasses}},\ }\href
  {https://doi.org/10.21468/SciPostPhys.1.2.016} {\bibfield  {journal}
  {\bibinfo  {journal} {SciPost Phys.}\ }\textbf {\bibinfo {volume} {1}},\
  \bibinfo {pages} {016} (\bibinfo {year} {2016})}\BibitemShut {NoStop}%
\bibitem [{\citenamefont {Lerner}\ \emph {et~al.}(2014)\citenamefont {Lerner},
  \citenamefont {DeGiuli}, \citenamefont {During},\ and\ \citenamefont
  {Wyart}}]{breakdown}%
  \BibitemOpen
  \bibfield  {author} {\bibinfo {author} {\bibfnamefont {E.}~\bibnamefont
  {Lerner}}, \bibinfo {author} {\bibfnamefont {E.}~\bibnamefont {DeGiuli}},
  \bibinfo {author} {\bibfnamefont {G.}~\bibnamefont {During}},\ and\ \bibinfo
  {author} {\bibfnamefont {M.}~\bibnamefont {Wyart}},\ }\bibfield  {title}
  {\bibinfo {title} {Breakdown of continuum elasticity in amorphous solids},\
  }\href {https://doi.org/10.1039/C4SM00311J} {\bibfield  {journal} {\bibinfo
  {journal} {Soft Matter}\ }\textbf {\bibinfo {volume} {10}},\ \bibinfo {pages}
  {5085} (\bibinfo {year} {2014})}\BibitemShut {NoStop}%
\bibitem [{\citenamefont {Liu}\ and\ \citenamefont {Nagel}(2010)}]{liu_review}%
  \BibitemOpen
  \bibfield  {author} {\bibinfo {author} {\bibfnamefont {A.~J.}\ \bibnamefont
  {Liu}}\ and\ \bibinfo {author} {\bibfnamefont {S.~R.}\ \bibnamefont
  {Nagel}},\ }\bibfield  {title} {\bibinfo {title} {The jamming transition and
  the marginally jammed solid},\ }\href
  {https://doi.org/10.1146/annurev-conmatphys-070909-104045} {\bibfield
  {journal} {\bibinfo  {journal} {Annu. Rev. Condens. Matter Phys.}\ }\textbf
  {\bibinfo {volume} {1}},\ \bibinfo {pages} {347} (\bibinfo {year}
  {2010})}\BibitemShut {NoStop}%
\bibitem [{\citenamefont {van Hecke}(2010)}]{van_hecke_review}%
  \BibitemOpen
  \bibfield  {author} {\bibinfo {author} {\bibfnamefont {M.}~\bibnamefont {van
  Hecke}},\ }\bibfield  {title} {\bibinfo {title} {Jamming of soft particles:
  geometry, mechanics, scaling and isostaticity},\ }\href
  {http://stacks.iop.org/0953-8984/22/i=3/a=033101} {\bibfield  {journal}
  {\bibinfo  {journal} {J. Phys.: Condens. Matter}\ }\textbf {\bibinfo {volume}
  {22}},\ \bibinfo {pages} {033101} (\bibinfo {year} {2010})}\BibitemShut
  {NoStop}%
\bibitem [{\citenamefont {Rens}\ and\ \citenamefont
  {Lerner}(2019)}]{robbie_auxetic}%
  \BibitemOpen
  \bibfield  {author} {\bibinfo {author} {\bibfnamefont {R.}~\bibnamefont
  {Rens}}\ and\ \bibinfo {author} {\bibfnamefont {E.}~\bibnamefont {Lerner}},\
  }\bibfield  {title} {\bibinfo {title} {Rigidity and auxeticity transitions in
  networks with strong bond-bending interactions},\ }\href
  {https://doi.org/10.1140/epje/i2019-11888-5} {\bibfield  {journal} {\bibinfo
  {journal} {The European Physical Journal E}\ }\textbf {\bibinfo {volume}
  {42}},\ \bibinfo {pages} {114} (\bibinfo {year} {2019})}\BibitemShut
  {NoStop}%
\bibitem [{\citenamefont {Paoluzzi}\ \emph {et~al.}(2019)\citenamefont
  {Paoluzzi}, \citenamefont {Angelani}, \citenamefont {Parisi},\ and\
  \citenamefont {Ruocco}}]{Paoluzzi_2019}%
  \BibitemOpen
  \bibfield  {author} {\bibinfo {author} {\bibfnamefont {M.}~\bibnamefont
  {Paoluzzi}}, \bibinfo {author} {\bibfnamefont {L.}~\bibnamefont {Angelani}},
  \bibinfo {author} {\bibfnamefont {G.}~\bibnamefont {Parisi}},\ and\ \bibinfo
  {author} {\bibfnamefont {G.}~\bibnamefont {Ruocco}},\ }\bibfield  {title}
  {\bibinfo {title} {Relation between heterogeneous frozen regions in
  supercooled liquids and non-debye spectrum in the corresponding glasses},\
  }\href {https://doi.org/10.1103/PhysRevLett.123.155502} {\bibfield  {journal}
  {\bibinfo  {journal} {Phys. Rev. Lett.}\ }\textbf {\bibinfo {volume} {123}},\
  \bibinfo {pages} {155502} (\bibinfo {year} {2019})}\BibitemShut {NoStop}%
\bibitem [{\citenamefont {Lerner}\ \emph {et~al.}(2013)\citenamefont {Lerner},
  \citenamefont {During},\ and\ \citenamefont {Wyart}}]{nonlinear_excitations}%
  \BibitemOpen
  \bibfield  {author} {\bibinfo {author} {\bibfnamefont {E.}~\bibnamefont
  {Lerner}}, \bibinfo {author} {\bibfnamefont {G.}~\bibnamefont {During}},\
  and\ \bibinfo {author} {\bibfnamefont {M.}~\bibnamefont {Wyart}},\ }\bibfield
   {title} {\bibinfo {title} {Low-energy non-linear excitations in sphere
  packings},\ }\href {https://doi.org/10.1039/C3SM50515D} {\bibfield  {journal}
  {\bibinfo  {journal} {Soft Matter}\ }\textbf {\bibinfo {volume} {9}},\
  \bibinfo {pages} {8252} (\bibinfo {year} {2013})}\BibitemShut {NoStop}%
\end{thebibliography}

%

\end{document}